\def\be{\begin{equation}}
\def\ee{\end{equation}}
\def\bea{\begin{eqnarray}}
\def\eea{\end{eqnarray}}
\def\noin{\noindent}
\def\V{{\bf{V}}}

\def\x{{\bf{x}}}

\def\k{{\bf{k}}}
\def\p{{\bf{p}}}
\def\q{{\bf{q}}}
\def\r{{\bf{r}}}

\def\op{\omega_p}

\documentstyle[psfig,prl,aps]{revtex}
\begin{document}
\bibliographystyle{/cmcl3.b/research/jwang/Revtex/prsty}
\draft
\tighten
% The following line is crucial to get two-column format right
\twocolumn[\hsize\textwidth\columnwidth\hsize\csname @twocolumnfalse\endcsname
\title{Diffusion in a Random Velocity Field: Spectral Properties of a Non-Hermitian Fokker-Planck Operator}
\author{ J.~T.~Chalker and  Z.~Jane Wang$^{\dagger}$ }
\address{Theoretical Physics, University of Oxford,
1, Keble Road, Oxford, OX1 3NP, UK }
\date{\today}
\maketitle

\begin{abstract}
We study spectral properties of the Fokker-Planck operator
that describes particles diffusing in a quenched
random velocity field. This random
operator is non-Hermitian and has eigenvalues occupying a finite area 
in the complex plane. We calculate
the eigenvalue density and
averaged one-particle Green's function,
for weak disorder and dimension $d>2$. We relate our
results to the time-evolution of particle density,
and compare them with
numerical simulations.

\end{abstract}
\vskip2pc] \narrowtext

%\section{Introduction}
In contrast to closed quantum systems,
classical systems often have dynamics generated by non-Hermitian 
operators. In this paper we develop general techniques to study the spectral
properties of random non-Hermitian operators, 
and apply them to
the Fokker-Planck (FP) operator that describes
diffusion and advection
of classical particles
in a spatially random but time-independent velocity field:

\be
{\partial_t n} = {{\cal L}_{fp}} n \equiv D 
\nabla^2 n - {\nabla} \cdot ({\V} n),
\label{fp-eq}
\ee

\noindent 
where $n$ is the concentration of particles, 
$D$ the molecular diffusivity, and $\V$ the background
velocity field. The non-Hermitian character, in this case, is due to the
advection term.

The statistical behavior of the scalar field $n$ 
obeying this FP equation has been investigated over 
a long history \cite{general,sinai,fis84-85,kraich}, in the context 
of turbulent diffusion and anomolous diffusion in random media.
However, for general velocity distributions,
little appears to be known about
spectral properties of the FP operator, such as the density of
eigenvalues (DOS).
An exception is the special
case of potential flow, with $\V =\nabla \phi$.
In this instance,  there is a similarity
transformation which maps the FP equation exactly
to a Schr\"{o}dinger equation in imaginary time\cite{ris89}.
The eigenvalues of the FP operator with potential flow 
are therefore real and
negative.
In one dimension, it is possible to
express any velocity field in terms of a potential
and to transform the FP equation in this way.
Moreover, anomolous diffusion in one-dimensional
systems with random flow has been shown
to be connected with logarithmic singularities
of the DOS  and of the eigenstate localization
length as the eigenvalue approaches zero \cite{tos88}.
In two dimensions, in the opposite case of 
incompressible flow ($\nabla . \V =0$),
for which there is no similarity transformation
to a Hermitian operator, localization
of eigenfunctions of the FP operator
has been studied numerically \cite{mil96}.
In addition, an analogy between the classical FP
equation and the quantum random flux problem has
been analyzed \cite{mil96}. 

In general, the eigenvalues of a non-Hermitian
FP operator occupy a 
finite area in the complex plane,
rather than being restricted to the real-axis.
This fact,  despite the similarities
in other respects between the FP and 
the Schr\"odinger equations,
renders inapplicable 
 \cite{som88} the standard perturbation 
expansion of Green's functions, used for disordered quantum
problems.
Furthermore, saddle-point techniques  \cite{som88,leh92}
developed for 
non-Hermitian random matrix ensembles \cite{meh90,gir85}
are too specialized to be appropriate
for random FP operators.

Very recently there has been considerable interest in properties of
random non-Hermitian operators \cite{hat96,recent,efetov,zee},
with a range of motivations, including the
study of open quantum systems \cite{leh92} and
the motion of flux lines in superconductors\cite{hat96}.
Against this background, a better understanding
of their spectral properties and of calculational
methods is clearly desirable. 

In this paper we describe a general scheme, based on a
diagrammatic expansion, to compute the disorder-averaged
Green's functions and the DOS of random non-Hermitian operators.
Similar ideas have been proposed in the
context of random matrix theory
by Feinberg and Zee \cite{zee}.
We apply the technique to the FP operator
of Eq.(\ref{fp-eq}),
calculating the shape of the support of the DOS,
and the eigenvalue density itself.
We also compare our analytical results with numerical calculations.

The particular FP operator we consider has constant 
diffusivity $D$ and a quenched random velocity field $\V(\x)$. 
Note that this is the opposite limit to that in the model recently discussed
by Kraichnan \cite{kraich}, which has infinitely short time correlation
in velocities. Time-independent flows can be established 
in physical systems such as porous media\cite{fis84-85}.

We take the velocity field to be gaussian distributed, with zero mean, and variance 

\bea
\langle\V_\alpha(\k)\V_\beta(\k')\rangle
 &=& \Gamma_1 ( \delta_{\alpha\beta} - \frac{k_\alpha
k_\beta} {\k^2}) \delta(\k+\k')  \nonumber\\
&+& \Gamma_2 ( \frac{k_\alpha
k_\beta} {\k^2}) \delta(\k+\k'),
\label{eq-gamma}
\eea

\noin where $\V(\k) = (2\pi)^{-d} \int d^dx e^{-i\k.\x}\V(\x)$, 
angular brackets denote the ensemble average,
and $\Gamma_1$ and $\Gamma_2$ represent the strengths of the
transverse and longitudinal parts of the velocity field.
We take the spectrum of velocity fluctuations to have a
short-wavelength cut-off, $\Lambda$, and consider a system of volume
$\Omega$.
The special case of mixed flow,
$\Gamma_1 = \Gamma_2 = \Gamma$, leads to substantial simplifications:
for clarity of presentation, we
describe calculations only in this limit, but state results
for the general problem. 

At a complex frequency $\omega$, the dimensionless combination 
$\gamma(\omega) = (\Gamma/|\omega|^2)(|\omega|/D)^{d/2+1}$
is a measure of the disorder strength.
%It has a simple physical interpretation: 
%let $L_A(t) = (\Gamma t^2)^{1/(d+2)}$ and 
%$L_D(t) = (D t)^{1/2}$
%be the mean distances a particle travels, owing to
%advection and diffusion respectively, in
%a time $t=|\omega|^{-1}$; then $\gamma(\omega) = (L_A(t)/L_D(t))^{d+2}$.
The fact that, as $\omega \to 0$, $\gamma(\omega) \to 0$ for $d>2$
and  $\gamma(\omega) \to \infty$ for $d<2$, identifies $d=2$ as
the upper critical dimension \cite{fis84-85}.

Our aim is to study spectral properties of the FP
operator, ${\cal L}_{fp}$. 
We do so via the ensemble-averaged Green's function

\be
g(\omega) = \langle \frac{1}{\omega - {\cal L}_{fp}} \rangle .
\ee 

\noin Let  $\langle L_{\lambda}|$ and 
$|R_{\lambda}\rangle$
be left and right eigenvectors of  
${\cal L}_{fp}$ with eigenvalue $\lambda$,
and let $|\p\rangle$ denote a plane-wave basis state
with wavevector $\p$.
The ensemble-averaged spectral density

\be
C(\p,\omega) = \langle \sum_{\lambda}
\langle \p |R_{\lambda} \rangle \langle L_{\lambda} | \p \rangle
\delta(\omega - \lambda)\rangle 
\ee

\noin is diagonal in this basis, because
averaging restores translational invariance.
From it we can obtain the
time Green's function, or particle density, $\langle n(\r,t)\rangle$,
evolving according to Eq (\ref{fp-eq}) with initial
condition $n(\r,t=0)=\delta(\r)$:

\be
\langle n(\r,t) \rangle =
 (2\pi)^{-d}\int d^d\p\,e^{i\p.\r} \int d^2\omega\,
e^{\omega t} C(\p,\omega)\,.
\label{n(r,t)}
\ee

\noin In the same basis, the diagonal elements of $g(\omega)$,
which we compute, are related to $C(\p,\omega)$ by

\be
g^p(\omega) = \int d^2 \lambda \frac{C(\p,\lambda)}{\omega - \lambda}\,.
\ee

Analytic properties of the Green's function depend
on the eigenvalue density, 
$\rho(\lambda)$, in the
complex $\lambda$-plane. In particular,

\be
\rho(\omega) = 
\frac{1}{\pi }\frac{\partial}
{\partial \omega^*}\frac{1}{\Omega} Tr g(\omega)\,.
\label{eq-rho}
\ee

\noin 
Thus, in the complex $\omega$-plane,
$g(\omega)$ is non-analytic everywhere 
that the eigenvalue density is non-zero. 
Standard techniques for calculating disorder-averaged
Green's functions via perturbation theory yield only
the part of $g(\omega)$ that is analytic in $\omega$, together with
its analytic continuation inside the support of $\rho(\omega)$
\cite{som88}. 
In the special case of pure potential flow,
$(\Gamma_1 = 0)$, the eigenvalues of ${\cal L}_{fp}$
all lie on the negative real axis
and  $g(\omega)$ is analytic elsewhere.
Under these circumstances one can compute, for example,
$\rho(\omega)$,
in the usual way,
from the discontinuity in  $g(\omega)$
across the real axis.
By contrast, for mixed flow (as we shall show), the eigenvalues of 
${\cal L}_{fp}$
fill a finite area in the complex plane,
the analytic part of $g(\omega)$
contains limited information, and a new approach is required.

To this end, for a general ${\cal L}$, we construct an matrix $\cal{H}$
which: (i) has twice the dimension of  ${\cal L}$; 
(ii) is Hermitian; and (iii) has an inverse that 
contains $g(\omega)$ as one of four blocks.
Specifically, with $A \equiv \omega - \cal{L}$,

\be
\cal{H} =\left( 
\begin{array}{cc}
\epsilon \hspace{0.5cm} A\\
A^{\dagger} \hspace{0.5cm} -1\\
\end{array}
\right).
\ee

\noin The inverse, $G \equiv {\cal H}^{-1}$, exists for real $\epsilon > 0$ 
and is

\bea
G &\equiv &\left( 
\begin{array}{c}
G_{11} \hspace{0.5cm} G_{12}\\
G_{21} \hspace{0.5cm} G_{22}\\
\end{array}
\right) \nonumber \\
&=&
\left( 
\begin{array}{c}
(\epsilon + A A^{\dagger})^{-1} \hspace{0.5cm} 
A(\epsilon + A^{\dagger}A)^{-1}\\
A^{\dagger}(\epsilon + A A^{\dagger})^{-1} \hspace{0.5cm}  
-\epsilon(\epsilon + A^{\dagger}A)^{-1}\\
\end{array}
\right).
\eea

\noin Since ${\cal H}$ is Hermitian, $G$ can be 
calculated using established methods, and from it we can obtain

\be
g(\omega) = \lim_{\epsilon \rightarrow 0} \langle G_{21} \rangle.
\ee

This approach parallels recent work by Feinberg and Zee \cite{zee}
and is somewhat different from that taken in
other calculations on spectral properties of non-Hermitian operators:
in the present notation, Sommers and collaborators \cite{som88,recent} 
focus on 
$\det(G_{11})$, while Efetov \cite{efetov}
separates ${\cal L}$ into Hermitian and anti-Hermitian parts.

A diagrammatic expansion for $\langle G \rangle$, and hence $g(\omega)$, 
follows from writing

\be
{\cal H} = {\cal H}_0 + {\cal H}_1 \ ,
\ee

\noin where

\be
{\cal H}_0 =\left( 
\begin{array}{cc}
\epsilon \hspace{0.5cm} 0\\
0 \hspace{0.5cm} -1\\
\end{array}
\right) \hspace{0.2cm} \mbox{and} \hspace{0.2cm}
{\cal H}_1 =\left( 
\begin{array}{cc}
0 \hspace{0.5cm} A\\
A^{\dagger} \hspace{0.5cm} 0\\
\end{array}
\right). 
\ee

\noin The series for $G$, in powers of ${\cal H}_1$ 
and of $G^{(0)} \equiv {\cal H}_0^{-1}$, involves two propagators, $G^{(0)}_{11}$ and  $G^{(0)}_{22}$,
and two vertices, $A$ and $A^{\dagger}$.
As usual, it is convenient to 
introduce a self-energy, $\Sigma$, and proceed via the Dyson equation,

\be
\langle G \rangle = G_0 + G_0 \Sigma \langle G \rangle,
\label{eq-dyson}
\ee

\noin where
\be
G_0=\left( 
\begin{array}{c}
1/\epsilon  \hspace{0.5cm} 0\\
0             \hspace{0.5cm} -1
\end{array}
\right).
\ee

To illustrate the approach, we consider
first the asymmetric gaussian random matrix ensemble of Ref \cite{som88}:
the $N \times N$ real matrix $[J]$ has the distribution
$P[J] \propto exp(-N Tr(JJ^{T}-\tau JJ)/[2(1-\tau^2)])$ 
so that non-zero covariances are

\be
<J^{T}_{ik}J_{ki}> = 1/N, \hspace{0.5cm} <J_{ik}J_{ki}> = \tau/N.
\ee

\noin The fully asymmetric problem, in which $\tau=1$ and
$J_{ik}$ and $J_{ki}$ are statistically independent,
was first studied by Ginibre \cite{meh90}, and has been
treated using a Green's function methods in Ref \cite{zee}.

Setting ${\cal L}=J$, the leading contribution to the self energy 
at large $N$ comes from the diagrams of the self-consistent Born
approximation (SCBA), shown in Fig \ref{fig-sigma}.

\begin{figure}[htb]
\centerline{\psfig{figure=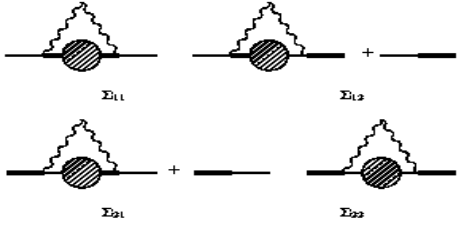,height=1.5in,width=3.0in}}
\caption{ Self-energy diagrams in the SCBA. Single and double
lines denote the two components of the bare $G_0$: $1/\epsilon$ and $-1$,
respectively, while the internal digrams describe the four full
Green's functions.}
\label{fig-sigma}
\end{figure}

\noin From these we obtain
\be
\Sigma =\left( 
\begin{array}{c}
\langle G_{22} \rangle  \hspace{0.5cm} 
\tau \langle G_{21} \rangle - \omega \\
\tau \langle G_{12}\rangle -\omega^{*} \hspace{0.5cm} 
\langle G_{11} \rangle \\
\end{array}
\right).
\ee

\noin Solving eq.(\ref{eq-dyson}), 
we recover the results of  Ref \cite{som88}.
In particular, inside the support of the
$\rho(\omega)$, defined by
$[x/(1+\tau)]^2+[y/(1-\tau)]^2 < 1$, we find

\be
N^{-1} Tr g(\omega) = ( \frac{x}{1+\tau} -\frac{i y}{1-\tau}), \hspace{0.5cm}
\omega = x + iy,
\ee

\noin and hence a constant DOS,
$\rho(\omega)=[\pi(1-\tau^2)]^{-1}$.

With this preparation, we return to the Fokker-Planck operator
of Eqs (\ref{fp-eq}) and
(\ref{eq-gamma}), taking 
$\Gamma_1=\Gamma_2 \equiv \Gamma$.
We find, at weak disorder and for dimension $d>2$,
that the self-energy is again given by the
SCBA. Corrections are small
by a factor of the dimensionless
disorder strength, $\gamma(x)$; 
neglecting these,
we obtain the self-energy (diagonal in wavevector, $p$)
\be
\Sigma_p =\left( 
\begin{array}{c}
p^2\Gamma\int d^d\q \langle G^{q}_{22} \rangle 
\hspace{2cm} -\omega_p \\
-\omega_p^{*} \hspace{2cm} \Gamma
\int d^d\q q^2 \langle G^{q}_{11} \rangle \\
\end{array}
\right),
\label{Sigma}
\ee

\noin where $\omega_p = \omega + D p^2$.

Before discussing the self-consistent 
solution to Eqs (\ref{eq-dyson}) and (\ref{Sigma}),
we note that one can arrive at the same point by
expressing $G$ in terms of a functional integral,
averaging using replicas,
and making a simple decoupling approximation.
In view of the $2 \times 2$ block structure of $G$,
it is natural to introduce two complex fields,
$\eta_1$ and $\eta_2$, with
$\eta^{\dagger}=(\eta_1^{\dagger},\eta_2^{\dagger})$,
and write for $j,k=1$ or $2$

\be
G_{jk} = i \langle Z^{-1} \int {\cal D} [\eta] 
\eta_j \eta_k e^{-iS} \rangle\,,
\ee

\noin where the normalisation and action are

\be 
Z =  \int {\cal D} [\eta] e^{-iS}
\hspace{0.2cm}
\mbox{and}
\hspace{0.2cm}
S = \eta^{\dagger} {\cal H} \eta \ .
\ee

\noin The average over the velocity field, denoted by $\langle \ldots
\rangle$ generates terms in the action of the form
$\eta_{1,\alpha}^{\dagger}
\eta_{1,\beta}
\eta_{2,\beta}^{\dagger}
\eta_{2,\alpha}$, where $\alpha$ and $\beta$ label replicas.
Approximating these by setting

\be
\eta^{\dagger}
\eta \,
\eta^{\dagger}
\eta
\approx
\langle
\eta^{\dagger}
\eta
\rangle
\eta^{\dagger}
\eta
+
\eta^{\dagger}
\eta
\langle
\eta^{\dagger}
\eta
\rangle
\ee

\noin we arrive at

\be
G_{ik} = i  \int {\cal D} [\eta] 
\eta_{i,\alpha} \eta_{k,\alpha} e^{-i{\tilde S}} ,
\ee

\noin with ${\tilde S} = [G^0]^{-1} - \Sigma$, where $\Sigma$ 
is again given by
Eq \ref{Sigma}.

Proceeding now to the evaluation of the self-energy, we find

\bea
\langle G_{11}^{p} \rangle &=& \frac{1 +  F_2}{|\omega_p|^2 + B_p}, \hspace{0.3cm}
\langle G_{12}^{p} \rangle = \frac{\op}{|\omega_p|^2 + B_p}
\nonumber\\
\langle G_{21}^{p} \rangle &=& \frac{\op^{*}}{|\omega_p|^2 + B_p},
\hspace{0.3cm}
\langle G_{22}^{p} \rangle =
 -\frac{\epsilon +p^2 F_1}{|\omega_p|^2 + B_p}
\,, \label{eq-sc}
\eea

\noin where $B_p=(\epsilon+F_1)(1+F_2)$ and 

\bea
F_1 =-\Gamma \int_0^{\Lambda} d^d\q  
\langle G_{22}^{q} \rangle , \nonumber \hspace{0.3cm} 
F_2 = \Gamma\int_0^{\Lambda} d^d\q  
\, q^2\langle G_{11}^{q} \rangle .\nonumber 
\eea

\noin Analysing this system of equations in the limit
$\epsilon \to 0$, we obtain 

\be
g^{p}(\omega) =  \frac{\omega_p^{*}}{|\omega_p|^2 + b p^2} \label{eq-g}\, .
\ee

\noin where the value of $b$ is determined via behaviour of the
integral

\be
I(b) = \Gamma \int_0^{\Lambda} d^d\q \frac{q^2}{|\omega_q|^2 + bq^2}.
\label{eq-I1}
\ee

\noin If $I(b=0)<1$, then $b=0$; otherwise $b$ is the 
(real and positive) solution to 
the equation $I(b)=1$. The former is the case if $\omega$ 
is not close to the negative
real axis. In that event, with $b=0$, 
$g^{p}(\omega) = \omega_p^{-1}$ and 
$\partial g^{p}(\omega)/\partial \omega^* = 0$, so that $\rho(\omega)=0$.
Alternatively, if $\omega$ is close to the negative real axis,
$b>0$,
$g^p(\omega)$ is not analytic in $\omega$ and
$\rho(\omega) \not= 0$.

Thus the boundary
to the support of the density of states, $y_B(x)$, satisfies the equation

\be
I(b=0) 
\equiv\Gamma 
\int_0^{\Lambda} d^d\q \frac{q^2}{(x+D q^2)^2 + y_B^{2}}=1.
\label{eq-bd}
\ee

\noin For small $\Gamma$, $d>2$ and $x<0$, we find

\be
y_B(x) = \pm \frac{\pi S_d}{2}\gamma(x)|x| \propto \pm \Gamma |x|^{d/2} 
\label{yB}
\ee

\noin where 
$S_d$ is the surface area of a $d$-dimensional unit sphere.
The DOS in the region $x<0, |y|<y_B$ is 
\be
\rho(\omega) = \frac{D}{(2\pi)^{d+1}\Gamma |x|}\,.
\label{rho}
\ee

\noin and elsewhere is zero.
Thus the eigenvalues occupy a wedge-shaped region in
the complex plane, centered around the negative real axis.
The $x$-dependent width of this region can be
understood simply: if one assumes that it is proportional to $\Gamma$,
dimensional analysis implies
$y_B(x) \propto \gamma(x) |x|$. Similarly, the
$x$-dependence of $\rho(\omega)$ follows from the requirement that $\int dy \rho(x+iy)$ take the value it has in the disorder-free system.

Turning to the time-evolution of the particle density,
we find that the
time Green's funtion, defined in Eq (\ref{n(r,t)}), is
the Fourier transform of a product of two factors: 

\be
\langle n(\r,t)\rangle= \frac{1}{(2\pi)^{d}}\int d^d\p\, e^{i\p.\r}\,
e^{-Dp^2t}\,[\frac{\sin(\omega_0t)}{\omega_0t}]\,, 
\label{n}
\ee

\noin where $\omega_0 = y_B(x)$, evaluated at $x=-Dp^2$: $\omega_0 = (\pi S_d \Gamma q^d)/(2 D)$. The first factor, $e^{-Dp^2t}$,
is the familiar consequence of simple diffusion; the second factor, 
$\sin(\omega_0t)/\omega_0t$, arises from advection. 
At weak disorder, when the SCBA is a controlled approximation,
the advective factor differs significantly from $1$ only
where the diffusive
factor is small, so that $\langle r^2 \rangle \sim Dt$. 
By contrast, at strong disorder (when the SCBA is simply
a mean-field approximation), it is the 
advective factor that sets the width of
of the density profile at short times, and  
$\langle r^2 \rangle \sim (\Gamma t/D)^{2/d}$.
This time-dependence arises because
advection and diffusion contribute equally 
to particle motion in this regime.
%In terms of the advection and diffusion lengths,
%$L_A(t)$ and $L_D(t)$, introduced after Eq(\ref{eq-gamma}),
%$\langle r^2 \rangle$ can be understood as the scale at which
%$L_A(t)=L_D(t)$.

For general random flow, with $\Gamma_1 \not= \Gamma_2$, an
extension of our approach again yields
Eqs (\ref{yB}), (\ref{rho}) and (\ref{n}), 
but with $\Gamma_1$ replacing $\Gamma$.
Thus, in particular, for potential flow ($\Gamma_1=0$) we 
reproduce correctly the fact that all eigenvalues are real.

\begin{figure}[htb]
\centerline{\psfig{figure=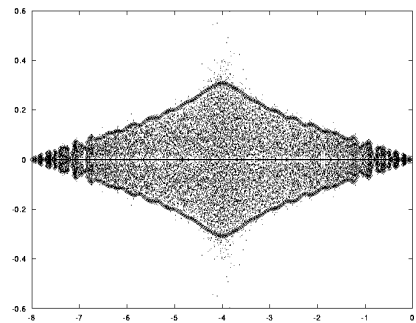,height=2in,width=2.5in}}
\caption{Distribution of eigenvalues (dots)
in the complex plane,  and calculated boundary ($\diamond$)
of $\rho(\omega)$.
 }
\label{fig-egv}
\end{figure}

To test the theory developed above, we have calculated numerically
the eigenvalues of the FP operator for mixed flow,
discretized on a square lattice. Theory (adapted to the
discretized FP operator) and simulation are compared in 
Fig (\ref{fig-egv}), for 50 realisations of a $32 \times 32$
lattice with $D=1$ and $\Gamma=0.25$.

\noin The calculated shape of the support of $\rho(\omega)$
matches the data well; the fact that a finite fraction of the eigenvalues 
have vanishing imaginary part is a finite-size effect,
analyzed in detail in Ref\cite{efetov}.

%In summary, we have presented a general technique to study
%spectral properties of random non-Hermitian operators,
%and an application of it to the FP operator describing 
%diffusion in a random velocity field.

J.T.C. is grateful to the Institute for Theoretical Physics, UCSB,
for hospitality during completion of this work, which was  
supported by EPSRC grant GR/GO 2727, and by NSF grant PHY94-07194; 
Z.J.W. acknowleges the support by an EPSRC grant, and
a NSF-NATO postdoctoral fellowship.

\noindent $\dagger$ Current address: 
Courant Institute of Mathematical Sciences,
New York University, 251 Mercer St., New York, NY 10012.
Electronic mail: jwang@cims.nyu.edu.

%\bibliography{/cmcl3.b/research/jwang/Paper/literature}

\begin{thebibliography}{10}

\bibitem{general}
For reviews, see: M.~B.~Isicheko,
Rev. Mod. Phys., {\bf 64}, 961 (1992), and
J.~P. Bouchaud and A.~Georges, Phys. Rep. {\bf 195}, 127 (1990).

\bibitem{sinai}
Ia. Sinai, in {\it Proceedings of the Berlin Conference in 
Mathematical  problems in theoretical Physics,} edited by
R. Schrader et. al. (Springer, Berlin, 1982), p12.

\bibitem{fis84-85}
D.~S. Fisher, Phys. Rev. A {\bf 30},  960  (1984);
D.~S. Fisher, Z. Qiu, S.~J. Shenker,
and S.~H. Shenker, {\it ibid.} {\bf 31},  3841  (1985);
J.~A. Aronovitz and D.~R. Nelson, {\it ibid.} {\bf 30},  1948  (1984);
V.~E. Kravtsov, I.~V. Lerner, and V.~I. Yudson, J. Phys. A{\bf 18}, L703 (1985);
Zh. Eksp. Teor. Fiz. {\bf 91}, 569 (1996) 
[Sov. Phys. JETP {\bf 64}, 336 (1986)]; 
Phys. Lett. A{\bf 119}, 203 (1986);
I.V. Lerner, Nucl. Phys. A {\bf 560},  274  (1993).

\bibitem{kraich}
R.~H.~Kraichnan, Phys. Rev. Lett. {\bf 72}, 1016 (1994).

\bibitem{ris89}
H. Risken, {\em The Fokker-Planck Equation} (Spring-Verlag, Berlin and
  Heidelberg, 1989).

\bibitem{tos88}
S. Alexander, J. Bernasconi, W.~R. Schneider, and R. Orbach,
Rev. Mod. Phys. {\bf 53}, 175 (1981);
E. Tosatti, M. Zannetti, and L. Pietronero, 
Zeit. Phys. B {\bf 73},  161  (1988).

\bibitem{mil96}
J. Miller and Z.~J. Wang, Phys. Rev. Lett. {\bf 76},  1461  (1996).

\bibitem{som88}
H.~J. Sommers, A. Crisanti, H. Sompolinsky, and Y. Stein, Phys. Rev. Lett. {\bf
  60},  1895  (1988).

\bibitem{leh92}
F. Hakke {\it et~al.}, Zeit. Phys. B {\bf 88},  359  (1992).
%N. Lehmann, D. Saher, V.~V. Sokolov, and H.~J. Sommers, Nucl. Phys. A {\bf
% 582},  223  (1995),

\bibitem{meh90}
J. Ginibre, J. Math. Phys. {\bf 6}, 440(1965),
M.~L. Mehta, {\em Random Matrices} (Academic Press Inc., Boston, 1990).

\bibitem{gir85}
V.~L. Girko, Theory Probab. Its Appl. (USSR) {\bf 29},  694  (1985).

\bibitem{hat96}
N. Hatano and D. R. Nelson, Phys. Rev. Lett. {\bf 77},  570  (1996).

\bibitem{recent}
Y.~V. Fyodorov, B. Khoruzhenko, and H.-J. Sommers, cond-mat/9606173,
Y.~V. Fyodorov, B. Khoruzhenko, and H.-J. Sommers, cond-mat/9702152.

\bibitem{efetov}
K.~B. Efetov, cond-mat/9702091.

\bibitem{zee}
J. Feinberg and A. Zee, hep-th/9703087
\end{thebibliography}
%\end{document}

\end{document}